\documentclass[%
  reprint,
  superscriptaddress,
  amsmath,amssymb,
  aps,
  prl,
  floatfix,
]{revtex4-2}

\usepackage{graphicx}
\usepackage{dcolumn}
\usepackage{bm}
\usepackage{hyperref}
\usepackage{xcolor}

\newcommand{\pourcentwidth}{0.95}                               
\newcommand{\nt}[1]           {#1}                              

\newcommand{\vc}  {V_{\mathrm{ch}}}
\newcommand{\vg}  {V_{\mathrm{g}}}

\newcommand{\eps} {\varepsilon}

\newcommand{\op}[1] {\mathbf{#1}} 

\newcommand{\abs}[1]    { {\left\vert #1 \right\vert}}

\newcommand{\GF}[2] {\langle\!\langle\, {#1}; \, {#2}\,\rangle\!\rangle}

\begin{document}

\title{\nt{Kondo-assisted switching between three conduction states in capacitively coupled quantum dots}}

\author{Pierre Lombardo}\email{pierre.lombardo@univ-amu.fr}
\affiliation{Aix-Marseille University, Faculty of Science, St J\'er\^ome, Marseille, France}
\affiliation{Institut Mat\'eriaux Micro\'electronique Nanosciences de Provence, UMR CNRS 7334, Marseille, France}

\author{Roland Hayn}
\affiliation{Aix-Marseille University, Faculty of Science, St J\'er\^ome, Marseille, France}
\affiliation{Institut Mat\'eriaux Micro\'electronique Nanosciences de Provence, UMR CNRS 7334, Marseille, France}

\author{Denis Zhuravel}
\affiliation{Bogolyubov Institute for Theoretical Physics, Kyiv 03143, Ukraine}

\author{Steffen Sch\"afer}
\affiliation{Aix-Marseille University, Faculty of Science, St J\'er\^ome, Marseille, France}
\affiliation{Institut Mat\'eriaux Micro\'electronique Nanosciences de Provence, UMR CNRS 7334, Marseille, France}

\date{June 30, 2020}

\begin{abstract}
  We propose a nanoscale device consisting of a double quantum dot with strong intra- and interdot Coulomb repulsions.  In this design, the current can only flow through the lower dot, but is triggered by the gate-controlled occupancy of the upper dot.  At low temperatures,  our calculations predict the double dot to pass through a narrow Kondo regime, resulting in highly sensitive switching characteristics between three well-defined states -- insulating, normal conduction and resonant conduction.
\end{abstract}

\maketitle



Soon after the advent of the first single-electron transistors in the late 1980s,\cite{AL1986,FD1987} with differential conductance plots governed by the hallmark {\sl Coulomb diamonds} due to charging electrons one by one to the central quantum island,\cite{LPWEUD1991,Kastner1992} it became clear that the Kondo effect\cite{Hewson1993book} would present a viable route to overcome the Coulomb blockade and to restore optimal conductivity.  Kondo physics has since been observed in a large variety of nanoscale devices, ranging from the original GaAs/AlGaAs\cite{GSM1998} and Si/SiGe\cite{KSE2007}  heterostructures to more exotic devices involving single-molecule junctions\cite{LSB2002,SN2010,MPHP2017,YN2004} or carbon nanotubes.\cite{NCL2000}  The sharpness of the concomitant zero-bias or Abrikosov-Suhl-Kondo resonance in the differential conductance,\cite{GSM1998,KSE2007,LSB2002,WFF2000,NCL2000,KMH2013,DRA2018} offers an attractive way to implement highly sensitive switching properties on which e.g. future molecular electronics might depend.\cite{MF2018,LTS2019}  

Like other quantum phenomena, the resonant spin-flip correlations underlying the Kondo effect are limited to very low temperatures and persist only up to some tens of milli-Kelvin in the usual semiconductor-based devices, or at best some tens of Kelvin in molecular electronics,\cite{MF2018,YN2004} although Kondo temperatures as high as $100\,\mathrm{K}$ have been reported for magnetic impurities on surfaces.\cite{WDS2004,TGN2019}  

\begin{figure}
\begin{center}
  \includegraphics[width=0.6\columnwidth]{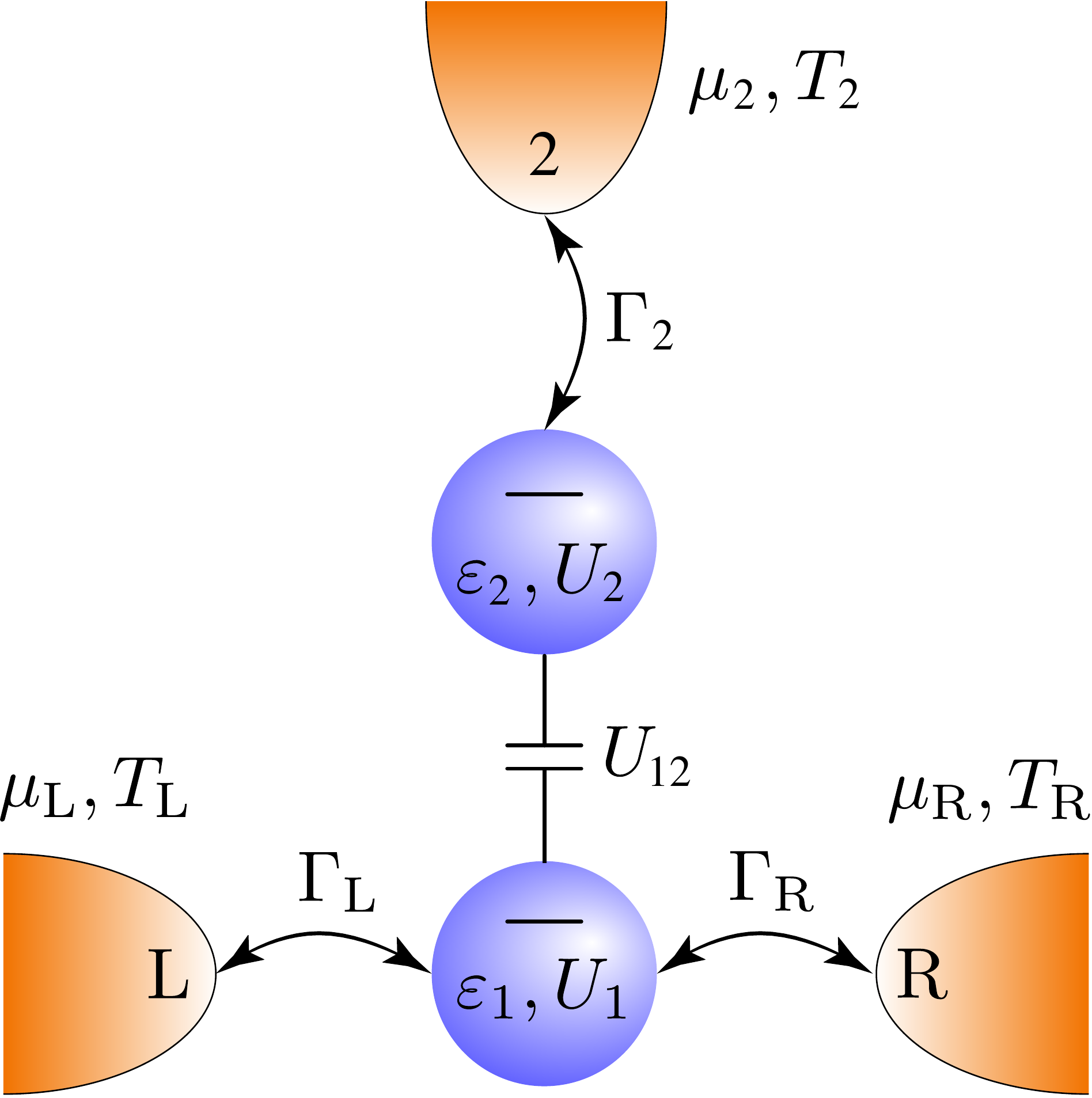}
  \caption{\label{fig0}
    Nanoscale \nt{device} where two single-level QDs, $\eps_1$ and $\eps_2$, with intradot Coulomb repulsion $U_1$ and $U_2$, and interdot Coulomb coupling $U_{12}$, are connected to three uncorrelated metallic leads.   A electron can tunnel from the left (L) lead via the lower dot to the right (R) lead, thus representing \nt{the channel of a transistor}.  The top lead controls the charge on the upper dot which, although isolated from the channel, triggers the current through the latter and thus acts as a gate. The tunneling between leads and their respective dots is summarized by the hybridization functions $\Gamma_{\alpha}$ where $\alpha\in\{L,R,2\}$.
    }
\end{center}
\end{figure}

The possibility to observe Kondo-assisted electronic transport in an otherwise Coulomb-blocked quantum dot (QD), hinges on the fact that a singly-occupied dot level is available for resonant tunneling from the leads, and is thus generally directly controlled via the gate voltage.\cite{GSM1998,KSE2007,MF2018}  For the $\bot$-shaped device proposed in this Letter by contrast, schematically depicted in Fig.~\ref{fig0}, the gate (lead 2) does not directly control the energy level $\eps_1$ of dot 1 in the conduction channel between the left (L) and right (R) lead, but addresses a second QD.  And it is the occupation of this upper dot, labeled 2, which ultimately triggers the onset of the current via an interdot Coulomb repulsion. \nt{As we will show in the last part of this Letter, the observed abruptness of the current onset is intimately linked to the presence of a Kondo feature in the spectral densities and to its exact location with respect to the Fermi level.}  Experimentally, capacitively coupled quantum dots without interdot tunneling have been realized e.g. in GaAs/AlGaAs heterostructures,\cite{CWM2002,HWDK2007} and tunable interdot couplings with minimal residual interdot tunneling were implemented using bilayer graphene on silicon substrate.\cite{FVT2012} While unarguably more complex, we will show in the following that, at low enough temperature, this double-dot setup offers the possibility to construct a quantum \nt{device} capable of switching in a highly sensitive manner between three states instead of two: insulating, normal conduction and Kondo-assisted resonant conduction.

%


To investigate the setup proposed in Fig.~\ref{fig0}, we will study the corresponding single-impurity Anderson model\cite{A1961} within the framework of the non-crossing approximation (NCA)\cite{MJP2000,OK2006,ZCR2013}.  The model's electronic Hamiltonian contains contributions from the leads, the dots, and from tunneling between leads and dots, $H=H_\mathrm{leads}+H_\mathrm{dots}+H_\mathrm{tun}$.  The three uncorrelated metallic leads, $\alpha\in\{L,R,2\}$, are described by
  \begin{subequations}
    \label{eq:Ham}
    \begin{equation}
      H_{\mathrm{leads}}\ =\ \sum_{\alpha k \sigma} \eps_{\alpha k}\,
      \op{c}_{\alpha k\sigma}^{\dagger}\, \op{c}_{\alpha k\sigma}  \,
      \text{,}
      \label{eq:Hleads}
    \end{equation}
    with $\op{c}_{\alpha k\sigma}^{\dagger}$ and $\op{c}_{\alpha k\sigma}$ the creation and annihilation operators of $\sigma$-electrons in state $k$ on the lead $\alpha$.  The contribution from the dots is
    \begin{equation}
      H_{\mathrm{dots}}\, =\,  \sum_{i \in \{1,2\}} \left(\eps_i\,\op{n}_i + U_i\,\op{n}_{i \uparrow} \op{n}_{i \downarrow}\right)
      \,+U_{12}\, \op{n}_1 \op{n}_2
      \text{,}
      \label{eq:Hdots}
    \end{equation}
    where $\op{n}_{i}=\op{n}_{i\uparrow}+\op{n}_{i\downarrow}$ is the total electronic occupation on dot $i$, with $\op{n}_{i\sigma}=\op{d}_{i\sigma}^{\dagger}\op{d}_{i\sigma}$ the number of $\sigma$-electrons on the dot, and $\op{d}_{i\sigma}^{\dagger}$ ($\op{d}_{i\sigma}$) the corresponding creation (annihilation) operators. The sum in the above expression thus describes two individual Anderson dots, with the upper dot serving as gate, $\vg\equiv\eps_2$, and the lower one defining the conduction channel, $\vc\equiv\eps_1$.   The last term in eq.~(\ref{eq:Hdots}) accounts for the capacitive coupling between both dots and can be implemented experimentally as described in Refs.~\onlinecite{CWM2002,HWDK2007,FVT2012}.

    Finally, the lead-dot tunneling is given by
    \begin{multline}
      H_{\mathrm{tun}}\, =\, \sum_{\substack{\alpha\in\{L,R\}\\k\sigma}}
      \left( t_{\alpha k}\, \op{d}^{\dagger}_{1\sigma}\,\op{c}_{\alpha k\sigma}
      \,+\, t^*_{\alpha k}\, \op{c}^{\dagger}_{\alpha k\sigma}\,\op{d}_{1\sigma} \right)
      \\
      + \sum_{k\sigma}
      \left( t_{2 k}\, \op{d}^{\dagger}_{2\sigma}\,\op{c}_{2 k\sigma}
      \,+\, t^*_{2 k}\, \op{c}^{\dagger}_{2 k\sigma}\,\op{d}_{2\sigma} \right)
      \,\text{,}
      \label{eq:Htun}
    \end{multline}
    where the first line describes the coupling of electrons on dot 1 to the left and right lead, while the second line allows electrons on dot 2 to tunnel to the upper lead.  
  \end{subequations}
  For most practical cases, the tunneling amplitudes $t_{\alpha k}$ only depend on $k$ via the corresponding level energy, $\eps_{\alpha k}$, such that the tunneling to each lead is entirely summarized by the hybridization strength\cite{JauhoWingreenMeir1994} $\Gamma_{\alpha}(\eps)= 2\pi \abs{t_\alpha(\eps)}^2 N_{\alpha}(\eps)$ where $N_{\alpha}(\eps)$ is the spin-summed DOS on lead $\alpha$ ($\in\{L,R,2\}$).
  Here, we further assume that all three leads are identical metals, with a DOS characterized by a single wide band, and that the tunneling between any lead and its dot is governed by the same only weakly energy dependent physical process such that all three hybridization strengths are the same, $\Gamma_{L}(\eps)=\Gamma_{R}(\eps)=\Gamma_{2}(\eps)$.  Specifically, we take broad Gaussian lead DOSes, of half-width $D=70\,\Gamma$, where $\Gamma=\Gamma_L(\bar{\mu})+\Gamma_R(\bar{\mu})$ is the total hybridization strength of the lower dot at the mean chemical potential $\bar{\mu}=\frac{1}{2}\left(\mu_L+\mu_R\right)$. $\Gamma$ will henceforth serve as our energy unit \nt{which, with $k_{\mathrm{B}}$ set to unity, will also be our unit of temperature.}   Furthermore, we assume zero voltage and temperature bias such that $\bar{\mu}=\mu_L=\mu_R=\mu_2$ and $T=T_L=T_R=T_2$.


  The conductance $G(T) = e^2 I_0(T)$ and Seebeck coefficient $S(T) = -I_1(T)/[e T I_0(T)]$ are given in terms of energy-weighted integrals,\cite{KH2003}
  \begin{equation}
    I_n(T)= \frac{2}{h}\int  \eps^n   \left( -\frac{\partial f}{\partial \eps} \right) \tau^{\rm eq}(\eps)     {\rm d}\eps\;\text{.}
  \end{equation}
  The integrand comprises the Fermi function, $f(\eps)$, and the equilibrium transfer function, $\tau^{\rm eq}(\eps)=\frac{\pi}{4} A_1(\eps)\Gamma(\eps)$.  Similarly, the occupancies of the dots $j=1,2$ also follow from the lead Fermi functions and the dot spectral functions $A_j(\eps)$ via\cite{WJM1993}
  \begin{displaymath}
    n_j=\int f(\eps-\mu_{j}) A_j(\eps){\rm d}\eps\,\text{.}  
  \end{displaymath}

  The $A_j(\eps)=-\frac{1}{\pi}{\rm Im}\sum_{\sigma}G^{\rm r}_{j\sigma}(\eps+i\delta)$  are readily obtained from the corresponding retarded dot Green's functions $G^{\rm r}_{j\sigma}(\eps+i\delta)=\GF{\op{d}_{j\sigma}}{\op{d}_{j\sigma}^{\dagger}}$.  In the atomic limit, the dot Green's functions would show sharp transitions between the 4 local eigenstates of each dot.  The hybridization with the leads and the interdot Coulomb repulsion, however, create correlations and fluctuations between these 16 local states, and we address the latter within the framework of the NCA, an approximation suitable for resonant spin-fluctuations underlying the Kondo effect which are expected in the parameter regime under investigation.


\begin{figure}
\begin{center}
  \includegraphics[width=\pourcentwidth\columnwidth]{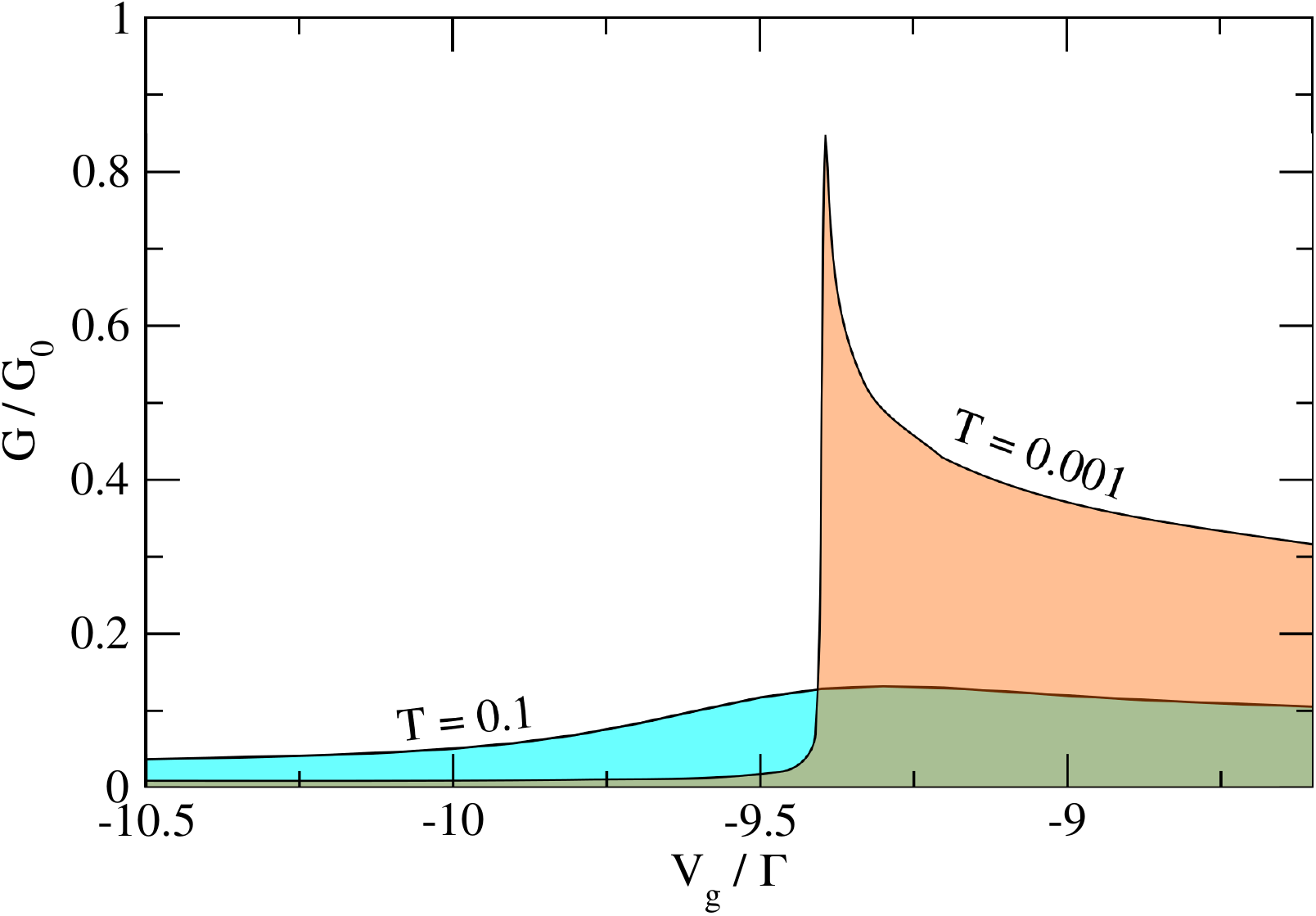}
  \caption{
    Conductance as a function of gate voltage $\vg\equiv\eps_2$ for high and low temperatures.  Parameters (in units of $\Gamma$) are $\vc\equiv\eps_1=-7.5$, intradot $U_1=U_2=8$ and interdot $U_{12}=6$.  At high temperatures, the conductance varies smoothly from insulating to metallic behavior.  At low temperatures, a sharp conductance resonance is observed at the transition from the insulating to the ordinary metallic state.
    \label{fig1}}
\end{center}
\end{figure}
%

\begin{figure}
\begin{center}
  \includegraphics[width=\pourcentwidth\columnwidth]{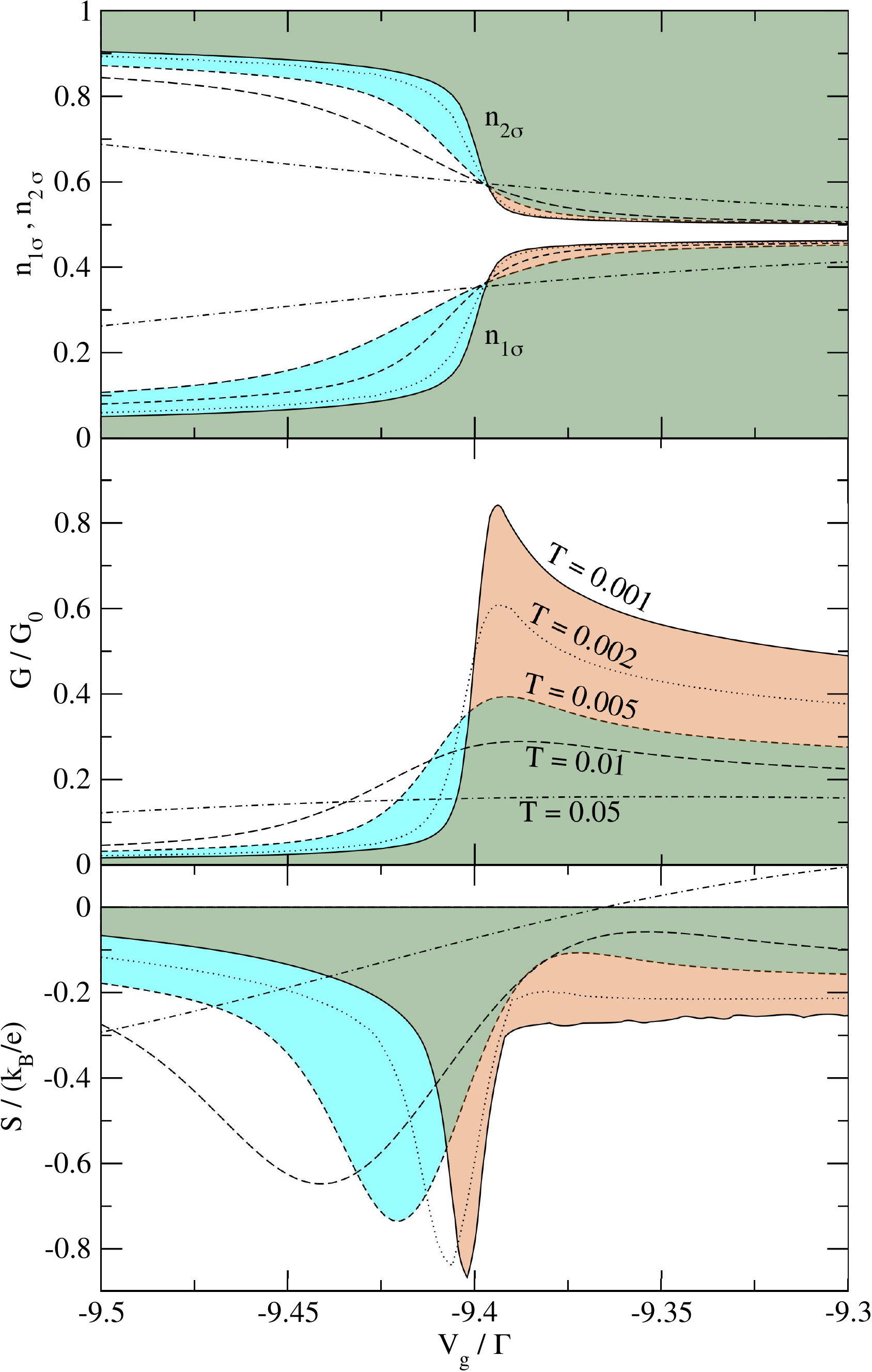}
  \caption{
    Dot occupancy per spin (top panel), conductance (middle), and Seebeck coefficient or thermopower (lower panel) for the same parameters as in Fig.~\ref{fig1} as a function of $\vg\equiv\eps_2$ close to the transition and for various temperatures.  Note that the total occupancy is twice the $n_{j\sigma}$ displayed in the upper panel, both for the lower dot (1), and the upper dot (2).
\label{fig2bis}}
\end{center}
\end{figure}

In our setup, we adopt a rather large intradot Coulomb repulsion, $U_1=U_2=8$, and an appreciable but not too large capacitive interdot coupling, $U_{12}=6$.  The energy level of the lower dot in the channel is fixed at $\vc\equiv\eps_1=-7.5$, while the upper dot level, $\vg\equiv\eps_2$, varies.  The resulting overall conductance is illustrated in Fig.~\ref{fig1}.  At high temperature, it shows a smooth crossover between a badly insulating and a badly conducting behavior.  At low temperatures, the  insulating and the metallic state are much closer to their defining characteristics, and the transition is very sharp.  Moreover, an additional narrow regime of enhanced conductance exists right above the transition voltage, $\vg^{\mathrm{c}}\approx -9.4$, in which $G$ is found to approach its theoretical maximum given by the conductance quantum {$G_0=\frac{2 e^2}{h}$}.  Note that this scenario is generic in the sense that it does not hinge on our specific choice of parameters: \nt{different Coulomb repulsions and hybridizations naturally shift the value of $\vg^{\mathrm{c}}$, but the characteristic switching behavior between three conduction states remains robust.  Conversely, a complete removal of the upper dot produces a quite different scenario in which the channel is insulating by default, and conductance is observed only for specific values of the gate voltage yielding the hallmark {\sl Coulomb diamond} plots.}

These findings are examined in greater detail in Fig.~\ref{fig2bis}, displaying the dot occupancies, the conductance and the thermopower for the same parameters as in Fig.~\ref{fig1}, but a much narrower range of gate voltages near the transition between insulating and metallic behavior.  \nt{As is well known, the Kondo effect requires a (spin-summed) dot occupancy not too far from unity.  We therefore choose parameters such that the total occupancy of both dots is close to two. As obvious from the symmetry of the occupancies in the upper panel of  Fig.~\ref{fig2bis}, our setup relies on an effect analogous to the famous communicating vessels to ensure that $n_1$ passes through the relevant regime upon variation of $\vg$.} The upper dot is close to doubly occupied for $\vg\ll\vg^{\mathrm{c}}$, which forces the occupancy of the lower dot to be small, thus impeding charge transport through the channel.   For $\vg\gg\vg^{\mathrm{c}}$, both dots are close to simply occupied and an ordinary metallic state arises which, at low temperature, has a conductance of approximately $\frac{1}{3} G_0$. 
  For low temperatures, we furthermore observe a strongly enhanced conductance on the metallic side of the transition, $\vg\gtrsim -9.4$.  For our lowest temperature, $T=0.001$, the enhanced conductance exhibits a maximum close to $G_0$.  The figure also shows that the maximum is strongly eroded with rising temperature. This, and the fact that the enhanced conductance is only observed in a very narrow gate voltage regime, hints to resonant spin fluctuation underlying the Kondo effect as a plausible cause. For our system, we indeed find a Kondo temperature of $T_{\mathrm{K}} \approx 8\cdot 10^{-3}$, in perfect agreement with the observed erosion of the conductance maximum. This scenario is further corroborated by the thermopower displayed in the last panel of Fig.~\ref{fig2bis}, since a negative or sign-changing thermopower are other hallmarks of Kondo-mediated transport.~\cite{CZ2010}
  
%

\begin{figure}[thb]
\begin{center}
  \includegraphics[width=\pourcentwidth\columnwidth]{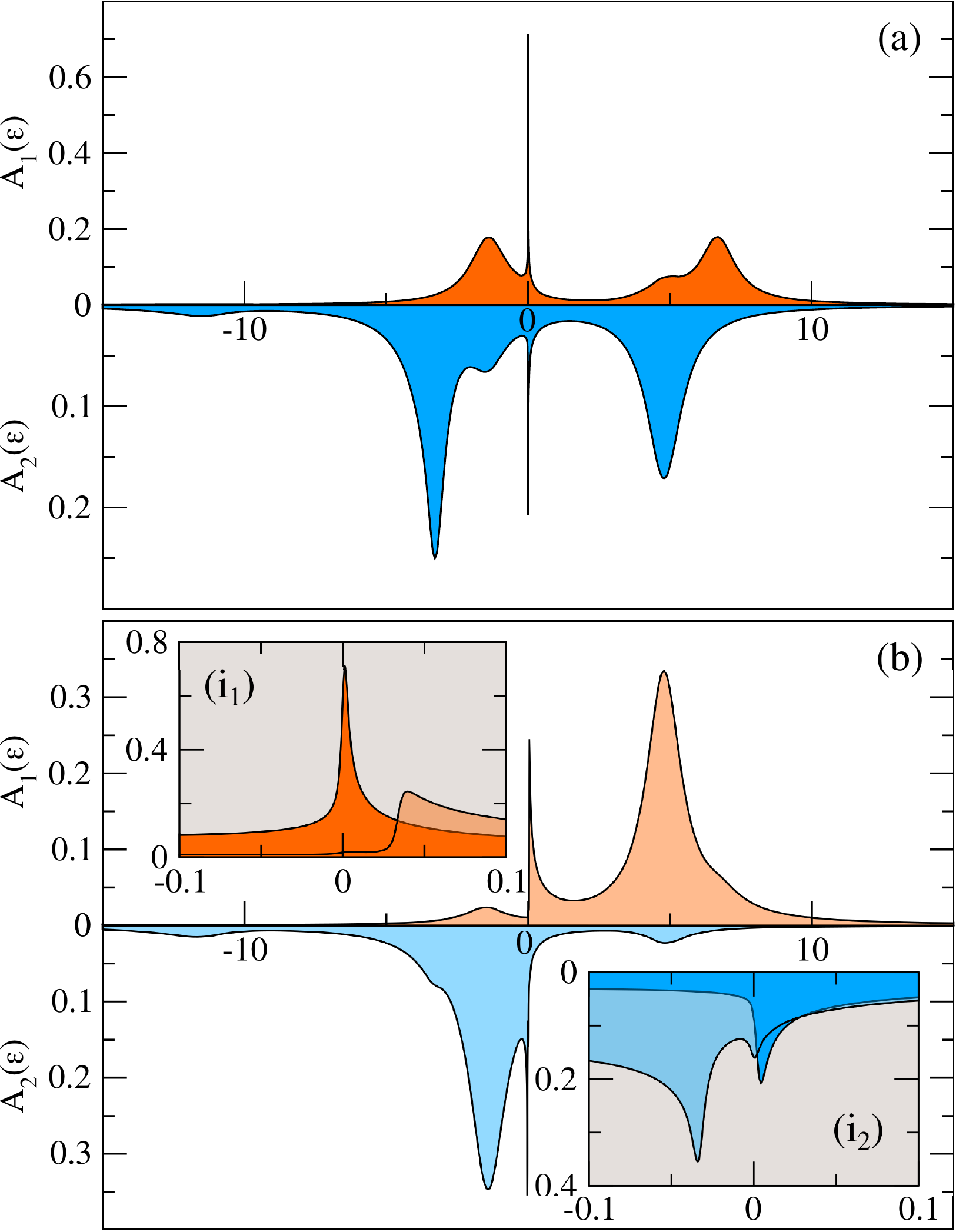}
    \caption{
      Spectral densities of the lower and upper dot, $A_1(\eps)$ and $A_2(\eps)$, with part (a) representing the metallic side, $\vg\approx\vg^{\mathrm{c}}+0.04$, and part (b) the insulating side, $\vg\approx\vg^{\mathrm{c}}-0.04$, of the transition, for the same parameters as in Fig.~\ref{fig1} and in the low-temperature regime, $T=0.001$.  The insets in the lower panel show blow ups of the lower ($\mathrm{i}_1$) and upper ($\mathrm{i}_2$) dot spectral functions around the Fermi level, with darker colors representing the metallic and lighter the insulating side of the transition.}
     \label{fig3bis}
\end{center}
\end{figure}

The low-temperature spectral densities displayed in Fig.~\ref{fig3bis} provide further support for a Kondo-enhanced conductance on the metallic side of the transition, and unveil the details behind the sudden switching behavior from virtually no to almost perfect conductance.  Part (a) shows the spectral densities of the lower dot (in red) and upper dot (in blue) on the metallic side of the transition, $\vg\approx\vg^{\mathrm{c}}+0.04$.  Both densities are dominated by their respective upper and lower Hubbard bands, separated by the intradot Coulomb $U_1=U_2=8$, as expected for roughly half-filled Anderson dots.  More importantly, a sharp Kondo resonance is present in both spectral densities exactly at the Fermi level (as is most obvious from the insets in the lower panel of the figure). It is thus the Kondo contribution to the spectral density at the Fermi level which allows to overcome the Coulomb blockade and to restore a perfectly metallic behavior -- an effect that has already been observed in other Kondo systems at low temperature.\cite{PPG2002}
  On the insulating side of the transition, $\vg\approx\vg^{\mathrm{c}}-0.04$, shown as part (b) of Fig.~\ref{fig3bis}, the spectral weight of the lower dot is predominantly transferred to the upper Hubbard band (in red), implying a small occupancy for this dot, whereas the spectral density of the upper dot resides mainly in its lower Hubbard band (in blue), indicating an almost fully occupied gate dot. Some Kondo contributions are still present close to the Fermi level, but they do not suffice to preserve the metallic character of the system:  as inset $\mathrm{i}_1$ shows most clearly, the residual Kondo peak (light red) lies too far above the Fermi level to contribute appreciably to the conductance through the dot.


  In summary, we have investigated the electronic transport through a double QD with strong intra- and interdot Coulomb repulsions in a $\bot$-shaped layout.  In this setup electrons may only flow from source to drain via the lower dot.  The role of the upper dot is to trigger the flow via an interdot Coulomb repulsion.  To achieve this, a gate regulates the occupancy of the upper dot.   A low temperatures, a very abrupt switching behavior is observed upon rising the gate voltage: first  insulating, the dot switches suddenly to a resonant regime where the conductance is close to the theoretical maximum of $G_0=2 e^2/h$, before adopting a normal metallic conductance.  Our calculations of the Seebeck coefficient and the spectral densities show that this peculiar switching behavior is intimately linked to the Kondo effect and its sharp zero-bias resonance.  Although limited to low temperatures, our three-state \nt{transistor-like device} might find applications in highly sensitive spectroscopy or open the road to ternary electronics.   

\bibliographystyle{apsrev4-2}

%

\end{document}